\documentclass[12pt]{article}

\begin{document}

\baselineskip 0.72cm
\topmargin -0.6in
\oddsidemargin -0.1in

\let\ni=\noindent

\renewcommand{\thefootnote}{\fnsymbol{footnote}}

\newcommand{\CKM}{Cabibbo--Kobayashi--Maskawa }

\newcommand{\SM}{Standard Model }

\pagestyle {plain}

\setcounter{page}{1}

\pagestyle{empty}


~~~~~
\begin{flushright}
IFT-01/26
\end{flushright}

{\large\centerline{\bf Fermion universality manifesting itself in}}
{\large\centerline{\bf the Dirac component of neutrino mass 
matrix{\footnote {Supported in part by the Polish State Committee for 
Scientific Research (KBN), grant 5 P03B 119 20 (2001--2002).}}}}

\vspace{0.3cm}

{\centerline {\sc Wojciech Kr\'{o}likowski}}

\vspace{0.2cm}

{\centerline {\it Institute of Theoretical Physics, Warsaw University}}

{\centerline {\it Ho\.{z}a 69,~~PL--00--681 Warszawa, ~Poland}}

\vspace{0.3cm}

{\centerline{\bf Abstract}}

\vspace{0.2cm}

\baselineskip 0.55cm

An effective texture is presented for six Majorana conventional neutrinos 
(three active and three sterile), based on a $6\times 6$ neutrino mixing 
matrix whose $3\times 3$ active--active component arises from the popular 
bimaximal mixing matrix of active neutrinos 
$ \nu_e\,, \,\nu_\mu\,, \,\nu_\tau $ by three small rotations in the 
14 , 25 , 36 planes of $ \nu_1\,, \,\nu_2\,, \,\nu_3 $ and 
$ \nu_4\,, \,\nu_5\,, \,\nu_6 $ neutrino mass states. The Dirac component 
({\it i.e.}, $3\times 3$ active--sterile component) of the resulting 
$6\times 6$ neutrino mass matrix is conjectured to get a structure 
{\it similar} to the charged--lepton and quark $3\times 3$ mass matrices, 
after the bimaximal mixing, specific for neutrinos, is {\it transformed out 
unitarily} from the neutrino mass matrix. The charged--lepton and quark mass 
matrices are taken in a {\it universal form} constructed previously by the 
author with a considerable phenomenological success. Then, for the option of 
$m^2_1\simeq m^2_2\simeq m^2_3\gg m^2_4\simeq m^2_5\simeq m^2_6\simeq0$, 
the proposed texture {\it predicts} oscillations of solar $\nu_e$'s with 
$\Delta m^2_{\rm sol} \equiv \Delta m^2_{21} \sim\, $(1.1 to 
1.2)$\,\times 10^{-5}\;{\rm eV}^2$, not inconsistent with the LMA solar 
solution, if the Super--Kamiokande value 
$\Delta m^2_{\rm atm} \equiv \Delta m^2_{32} \sim \,$(3 to 
3.5)$\,\times 10^{-3}\;{\rm eV}^2$ for oscillations of atmospheric 
$\nu_\mu$'s is taken as an input. Here, $\sin^2 2\theta_{\rm sol} \sim 1$ 
and $ \sin^2 2 \theta_{\rm atm} \sim 1$. The texture {\it predicts} also an 
LSND effect with $\sin^2 2\theta_{\rm LSND}\sim\,$(1.4 to 
1.9)$\,\times 10^{-11}\,({\rm eV}/m_1)^4$ and 
$\Delta m^2_{\rm LSND} \equiv \Delta m^2_{25} \sim m^2_1 + \,$(1.1 to 
1.2)$\,\times 10^{-5}\;{\rm eV}^2$. Unfortunately, the Chooz experiment 
imposes on the LSND effect (in our texture) a very small upper bound 
$\sin^2 2\theta_{\rm LSND} \stackrel{<}{\sim} 1.3 \times 10^{-3}$, 
which corresponds to the lower limit $ m_1 \stackrel{>}{\sim} \, $(1.0 
to 1.1)$\,\times 10^{-2}\,{\rm eV}$. 

\baselineskip 0.75cm

\vspace{0.1cm}

\ni PACS numbers: 12.15.Ff , 14.60.Pq , 12.15.Hh .

\vspace{0.6cm}

\ni September 2001

\vfill\eject

~~~
\pagestyle {plain}

\setcounter{page}{1}

\vspace{0.2cm}

\ni {\bf 1. Introduction}

\vspace{0.2cm}

Some time ago we constructed an effective form of fundamental--fermion 
mass matrix which worked very well for charged leptons 
$ e\,, \,\mu\,, \,\tau $ \cite{W1}, neatly for up and down quarks 
$ u\,, \,c\,, \,t $ and $ d\,, \,s\,, \,b $ \cite{W2} and badly for neutrinos 
$\nu_e\,, \,\nu_\mu\,, \,\nu_\tau $ \cite{W3}. Of course, if neutrinos 
are Majorana 
particles and/or, beside three active neutrinos carrying \SM charges, there 
exist effectively some sterile neutrinos free of these charges, it is natural 
to expect the neutrino texture to be different from charged--lepton and quark 
textures, characteristic for active Dirac particles. However, if neutrinos 
get a $3\times 3$ Dirac mass matrix $ M^{(D)}$ as a component of their 
overall $6\times 6$ mass matrix
 
\begin{equation}
M = \left( \begin{array}{cc} M^{(L)} & M^{(D)} \\ M^{(D)\dagger}  
& M^{(R)} \end{array} \right) \;,
\end{equation}

\ni then $M^{(D)}$ may display a similar structure as the charged--lepton 
and quark $3\times 3$ mass matrices.

Note that the neutrino mass matrix $M $ decribed in Eq. (1) leads to the 
neutrino mass term
 
\begin{equation} 
- {\cal L}_{\rm mass} = \frac{1}{2}\sum_{\alpha \beta} 
(\overline{\nu_\alpha^{(a)}} \,,\, \overline{\nu_{\alpha}^{(s)}}) 
\left( \begin{array}{cc} M^{(L)}_{\alpha \beta} & M^{(D)}_{\alpha \beta} \\ 
M^{(D)*}_{\beta \alpha}  & M^{(R)}_{\alpha \beta} \end{array} \right) 
\left( \begin{array}{c} \nu^{(a)}_\beta \\ \nu^{(s)}_{\beta} \end{array} 
\right) 
\end{equation}

\ni in the Lagrangian, where

\begin{equation} 
\nu^{(a)}_\alpha = \nu_{\alpha L} + 
\left( \nu_{\alpha L} \right)^c \,\;,\;\,\nu^{(s)}_\alpha 
= \nu_{\alpha R} + \left(\nu_{\alpha R} \right)^c \;,
\end{equation} 

\ni with $ \nu_\alpha = \nu_e\,,\,\nu_\mu\,,\,\nu_\tau \;\;(\alpha = 
e\,,\,\mu\,,\,\tau) $, are the Majorana conventional active and sterile 
neutrinos, respectively. Further on, for six neutrino flavor fields (states) 
we will use the notation $\nu^{(a)}_\alpha \equiv \nu_\alpha $ and 
$\nu^{(s)}_\alpha \equiv \nu_{\alpha_s}\;(\alpha = e\,,\,\mu\,,\,\tau $ and 
$ \alpha_s = e_s\,, \, \mu_s\,, \,\tau_s )$, and then pass to 
$ \nu_\alpha = \nu_e\,,\,\nu_\mu\,,\,\nu_\tau \,,\, \nu_{e_s} \,,\,  
\nu_{\mu_s} \,,\, \nu_{\tau_s} \;(\alpha = e\,,\,\mu\,,\,\tau \,,\,e_s\,, \, 
\mu_s\,, \,\tau_s )$. In the last case, six neutrino mass fields (states) 
will be denoted as $ \nu_i = \nu_1 \,,\, \nu_2 \,,\, \nu_3 \,,\, \nu_4 \,,\, 
\nu_5 \,,\, \nu_6 \;(i = 1,2,3,4,5,6)$. The relation between both 
complementary sets of neutrino fields (states) will read 
 
\begin{equation} 
\nu_\alpha  = \sum_i U_{\alpha i}  \nu_i \;,\;\left( |\nu_\alpha 
\rangle = \nu^\dagger_\alpha |0 \rangle =  \sum_i U^*_{\alpha i} |\nu_i 
\rangle \right) 
\end{equation} 

\ni with $ U = \left( U_{\alpha\,i} \right)$ being the unitary neutrino 
mixing matrix.

In the representation, where the charged--lepton $3\times 3$ mass matrix is 
diagonal, the neutrino mixing matrix $ U = \left( U_{\alpha\,i} \right)$ is 
at the same time the unitary matrix diagonalizing the neutrino mass matrix 
$ M = \left( M_{\alpha \,\beta} \right)$ given in Eq. (1):

\begin{equation} 
U^\dagger M U = M_{\rm d} \equiv 
{\rm diag}(m_1\,,\,m_2\,,\,m_3\,,\,m_4\,,\,m_5\,,\,m_6)\;.
\end{equation} 

\ni  Thus, in this case $ M = U M_{\rm d} U^\dagger $. 

The aim of the present paper is a brief discussion on the relevant question, 
to what extent the Dirac component $ M^{(D)}$ of the neutrino mass matrix 
$ M $ may get a {\it similar} structure to the charged--lepton and quark 
$ 3\times 3$ mass matrices constructed previously \cite{W1,W2}. In fact, 
in the 
cases of these active Dirac particles, we constructed with a considerable 
phenomenological success three $ 3 \times 3$ mass matrices of the 
{\it universal form}

\begin{equation}
M^{(D)} = \frac{1}{29} \left(\begin{array}{ccc} \mu \varepsilon & 2\alpha  
& 0 \\ & & \\ 2\alpha  & (4/9)\,\mu\, (80 + \varepsilon) & 
8\sqrt{3}\, \alpha \\ & & \\ 0 & 8\sqrt{3}\, \alpha &  
(24/25)\, \mu\, (624 + \varepsilon) \end{array} \right) \;,
\end{equation}

\ni where values of the constants $ \mu > 0 $, $\alpha > 0 $ and 
$\varepsilon > 0 $ depended on whether the discussed fermions were 
charged leptons or up quarks or down quarks (some foundations of our 
construction are collected in Appendix to the second Ref. \cite{W2}).

\vspace{0.2cm}

\ni {\bf 2. A model of neutrino texture}

\vspace{0.2cm}

In order to operate with a neutrino texture potentially consistent with 
oscillation experiments for solar $\nu_e$'s and atmospheric $\nu_\mu $'s 
as well as with LSND experiment for accelerator $\nu_\mu $'s, we will 
consider the $6\times 6$ neutrino mass matrix $ M = \left( 
M_{\alpha\,, \,\beta} \right)$  ($ \alpha \,\beta = 
e\,,\,\mu\,,\,\tau \,,\,e_s\,, \, \mu_s\,, \,\tau_s $) given in Eq. (1). We 
will assume that its diagonalizing matrix $ U = \left( U_{\alpha \,i} 
\right)$ ($ \alpha =  e\,,\,\mu \,,\,\tau \,,\,e_s\,, \, 
\mu_s\,, \,\tau_s \,,\, i = 1,2,3,4,5,6 $) can be written as

\begin{equation} 
U = {\stackrel{1}{U}}{\stackrel{0}{U}}\;,
\end{equation} 

\ni where the two factors are unitary matrices

\begin{equation} 
{\stackrel{1}{U}} =  \left( \begin{array}{cc} 
U^{(3)} & 0^{(3)} \\ 0^{(3)} & 1^{(3)} \end{array} \right) \;, \;
{\stackrel{0}{U}} = \left( \begin{array}{cc} 
C^{(3)} & S^{(3)} \\ -S^{(3)} & C^{(3)} \end{array} \right)  
\end{equation} 

\ni defined through the $3\times 3 $ submatrices

\begin{equation} 
U^{(3)} =   \left( \begin{array}{ccc} \frac{1}{\sqrt{2}} & 
\frac{1}{\sqrt{2}} & 0 \\ -\frac{1}{2} &  \frac{1}{2} & 
\frac{1}{\sqrt{2}} \\ \frac{1}{2} & -\frac{1}{2} & 
\frac{1}{\sqrt{2}}  \end{array} \right) \;,
1^{(3)} =  \left( \begin{array}{ccc} 1 & 0 & 0 \\ 
0 & 1 & 0 \\ 0 & 0 & 1 \end{array} \right) \;,\; 
0^{(3)} =  \left( \begin{array}{ccc} 0 & 0 & 0 \\ 
0 & 0 & 0 \\ 0 & 0 & 0 \end{array} \right) 
\end{equation} 

\ni and

\begin{equation} 
C^{(3)} =  \left( \begin{array}{ccc} c_{14} & 0 & 0 \\ 0 & c_{25} & 
0 \\ 0 & 0 & c_{36} \end{array} \right) \;,\;  S^{(3)} =  
\left( \begin{array}{ccc} s_{14} & 0 & 0 \\ 0 & s_{25} & 
0 \\ 0 & 0 & s_{36} \end{array} \right) \;,
\end{equation} 

\ni while $ c_{i j} = \cos \theta_{i j}$ and $ s_{i j} = \sin \theta_{i j}$ 
(the possible CP violating phases are ignored). Here, $ U^{(3)}$ involving 
$ c_{12} = 1/\sqrt{2} = s_{12} \,,\,c_{23} = 1/\sqrt{2} = s_{23} 
\,,\, c_{13} = 1 \,,\,s_{13} = 0$ and the phase $\delta = 0$ is the popular 
bimaximal mixing matrix for active neutrinos 
$\nu_e\,,\,\nu_\mu\,,\,\nu_\tau\, $, describing in a reasonable approximation 
oscillations of solar $\nu_e$'s and atmospheric $\nu_\mu $'s as suggested by 
SNO and Super--Kamiokande experiments \cite{W4,W5}. Using Eqs. (7) -- (10) 
we obtain explicitly

\begin{equation} 
U =\left( U_{\alpha i} \right) =  \left( \begin{array}{cccccc} 
\frac{c_{14}}{\sqrt{2}} & \frac{c_{25}}{\sqrt{2}} & 0 & 
\frac{s_{14}}{\sqrt{2}} & \frac{s_{25}}{\sqrt{2}} 
& 0  \\ -\frac{c_{14}}{2} & \frac{c_{25}}{2} & \frac{c_{36}}{\sqrt{2}} & 
-\frac{s_{14}}{2} & \frac{s_{25}}{2} & \frac{s_{36}}{\sqrt{2}} \\ 
\frac{c_{14}}{2} & -\frac{c_{25}}{2} & \frac{c_{36}}{\sqrt{2}} & 
\frac{s_{14}}{2} & -\frac{s_{25}}{2} & \frac{s_{36}}{\sqrt{2}}\\ 
-s_{14} & 0 & 0 & c_{14} & 0 & 0 \\ 0 & -s_{25} & 0 & 0 & c_{25} & 
0 \\ 0 & 0 & -s_{36} & 0 & 0 & c_{36} \end{array} \right) \;.
\end{equation}

\ni When $ s_{14}\,,\,s_{25}\,,\,s_{36}\rightarrow 0 $, then 
${\stackrel{0}{U}}\rightarrow 1$ and so, $U \rightarrow {\stackrel{1}{U}}$ 
and $ M = U M_{\rm d} U^\dagger \rightarrow {\stackrel{1}{U}} M_{\rm d} 
{\stackrel{1}{U}\!^\dagger}$. Here, ${\stackrel{1}{U}} M_{\rm d} 
{\stackrel{1}{U} \!^\dagger} = M_{\rm d}$ if the neutrino mass spectrum $m_i$ 
were degenerate for i = 1,2,3: $ m_1 = m_2 = m_3 $. Thus, in the above limit, 
it would be $ M \rightarrow M_{\rm d}$ in the case of such degeneracy of 
$ m_i$.

From Eqs. (5) and (7) we can deduce that

\begin{equation} 
{\stackrel{0}{U}\!^\dagger} {\stackrel{0}{M}}{\stackrel{0}{U}} = M_{\rm d} 
\equiv {\rm diag}(m_1\,,\,m_2\,,\,m_3\,,\,m_4\,,\,m_5\,,\,m_6)\;,
\end{equation}

\ni where ${\stackrel{0}{M}}$ is defined by the unitary transformation 
of $ M $, generated by ${\stackrel{1}{U}} $:

\begin{equation} 
{\stackrel{0}{M}} = {\stackrel{1}{U}\!^\dagger} M {\stackrel{1}{U}}\;.
\end{equation}
\ni Thus, ${\stackrel{0}{M}} = {\stackrel{0}{U}}M_{\rm d} 
{\stackrel{0}{U}\!^\dagger} $ since $M = U M_{\rm d} U^\dagger $. Further on, 
we will conjecture ({\it cf.} Eq. (19) later on), that the Dirac component 
of neutrino mass matrix ${\stackrel{0}{M}} = {\stackrel{1}{U}\!^\dagger} 
M {\stackrel{1}{U}}$ (rather than that of the full neutrino mass matrix 
$M $) gets a similar structure to the charged--lepton and quark $3\times 3$ 
mass matrices constructed previously \cite{W1,W2} in the universal form (6). 
If the Dirac component of the full neutrino mass matrix $ M $ were similar in 
structure to the charged--lepton and quark mass matrices of the universal 
form (6), then in the neutrino case (as is not difficult to show) there would 
be $ \mu = 0 $, $\alpha = 0 $ and $\varepsilon = 0 $ {\it i.e.}, the Dirac 
component of $ M $ would vanish trivially. In the accepted option, only 
parameter $\alpha $ must be zero for neutrinos. (Note that, in the spirit 
of fermion universality, the value $\alpha^{(\nu)} = 0 $ when correlated with 
the electric charge $ Q^{(\nu)} = 0 $ is consistent with our previous 
conjecture \cite{W2} that for up and down quarks 
$\alpha^{(u)} : \alpha^{(d)} = 
Q^{(u)\,2} : Q^{(d)\,2} = 4:1 $, where $Q^{(u)} = 2/3$ and $Q^{(d)} = -1/3$, 
and also with the fact that for charged leptons $\alpha^{(e)} \neq 0$, where 
$Q^{(e)} = -1 \neq 0$.) The appearance of $\stackrel{0}{M}$ (rather than 
$ M $) in the correct option for Dirac component of neutrino mass matrix 
may be understood qualitatively as related to the fact that the unitary 
transformation $ M \rightarrow \stackrel{0}{M}$ ({\it viz.} 
$\stackrel{0}{M} = \stackrel{1}{U} \!^\dagger M \stackrel{1}{U}$) 
eliminates from $ M $ the specific bimaximal mixing that makes neutrinos 
differ from other fundamental fermions (charged leptons and quarks). Thus, 
in our texture, the {\it fermion universality} manifests itself in the Dirac 
component of neutrino mass matrix {\it up to the unitary transformation} 
(of this matrix) {\it generated by $\stackrel{1}{U}$}. 

Note from Eq. (7) that ${\stackrel{1}{U}} =  
\left({\stackrel{1}{U}}_{\alpha i} \right) $ and 
${\stackrel{0}{U}} = \left({\stackrel{0}{U}}_{ij} \right) $, 
and so, ${\stackrel{0}{M}} = \left( {\stackrel{0}{M}}_{i j} \right)$ 
($i,j =1,2,3,4,5,6$) from Eq. (13). Due to the formula 
${\stackrel{0}{M}} = {\stackrel{0}{U}} M_{\rm d} 
{\stackrel{0}{U}}\,\!^\dagger$ we get explicitly

\begin{eqnarray}
\!\!\!{\stackrel{0}{M}}_{11} =  m_1 c^2_{14}+ m_4 s^2_{14} & , 
& {\stackrel{0}{M}}_{14} = {\stackrel{0}{M}}_{41} = 
(m_4-m_1) c_{14}s_{14}\;\;,\;\; {\stackrel{0}{M}}_{44} = 
m_1 s^2_{14}+m_4 c^2_{14}\;\;,\;\;\, \nonumber \\ 
\!\!\!{\stackrel{0}{M}}_{22} = m_2 c^2_{25}+m_5 s^2_{25} & , 
& {\stackrel{0}{M}}_{25} = 
{\stackrel{0}{M}}_{52} = (m_5-m_2) c_{25}s_{25}\;\;,\;\; 
{\stackrel{0}{M}}_{55} = m_2 s^2_{25}+m_5 c^2_{25}\;\;,\;\;\; 
\nonumber \\ \!\!\!{\stackrel{0}{M}}_{33} = m_3 c^2_{36}+m_6 s^2_{36} & , 
& {\stackrel{0}{M}}_{36} = {\stackrel{0}{M}}_{63} = 
(m_6-m_3) c_{36}s_{36} \;\;,\;\; {\stackrel{0}{M}}_{66} = 
m_3 s^2_{36}+m_6 c^2_{36}\;\;\;\;\;\;   
\end{eqnarray}

\ni and all other ${\stackrel{0}{M}}_{i j} = 0$. Notice the relations 
$ M_{e e_s} \!=\! {\stackrel{0}{M}}_{14}/ \sqrt{2}$, 
$ M_{\mu \mu_s} \!=\! {\stackrel{0}{M}}_{25}/2$, 
$ M_{\tau \tau_s} \!=\! {\stackrel{0}{M}}_{36}/\sqrt{2} $ between the 
diagonal elements of Dirac component $M^{(D)}$ of the full mass matrix 
$ M = \left( M_{\alpha \,\beta} \right)$ and the elements of Dirac component  
${\stackrel{0}{M}}\,\!^{(D)} \!=\! {\rm diag}({\stackrel{0}{M}}_{14} , 
{\stackrel{0}{M}}_{25} , {\stackrel{0}{M}}_{36}) $ of the mass matrix 
${\stackrel{0}{M}} \!=\! \left( {\stackrel{0}{M}}_{i j} \right) $. This is 
so, since $M^{(D)} = ({\stackrel{1}{U}} {\stackrel{0}{M}} 
{\stackrel{1}{U}}\,\!^\dagger )^{(D)} = U^{(3)}{\stackrel{0}{M}}\,\!^{(D)}$ 
with $ U^{(3)}$ given in Eq. (5) (similarly, $M^{(L)} = U^{(3)} 
{\stackrel{0}{M}}\,\!^{(L)} U^{(3)\,\dagger}$ and $M^{(R)} = 
{\stackrel{0}{M}}\,\!^{(R)} $).

From Eq. (14) we can infer that the neutrino masses $m_i$ can be expressed 
in the following way:

\begin{equation} 
m_{i,j} = {\stackrel{0}{M}}_{ii,jj} \mp \frac{s_{ij}}{c_{ij}} 
{\stackrel{0}{M}}_{i j} = {\stackrel{0}{M}}_{jj,ii} \mp 
\frac{c_{ij}}{s_{ij}}{\stackrel{0}{M}}_{i j} \;\; (ij = 14\,,\,25\,,\,36)\,,
\end{equation}

\ni if $ c_{ij} \neq 0$ and $ s_{ij} \neq 0$. These formulae imply the 
relations

\begin{equation} 
\frac{{\stackrel{0}{M}}_{ij}}{{\stackrel{0}{M}}_{jj} - 
{\stackrel{0}{M}}_{ii}} = \frac{c_{ij} s_{ij}}{c^2_{ij} - s^2_{ij}}\simeq 
\frac{s_{ij}}{c_{ij}}\;\; (ij = 14\,,\,25\,,\,36)\,,
\end{equation}

\ni where the last approximation is valid for $s^2_{ij} \ll c^2_{ij}$. In
 this approximation,

\begin{equation} 
m_{i,j} \simeq {\stackrel{0}{M}}_{ii,jj} \mp 
\frac{{\stackrel{0}{M}} \,\!^2_{ij}}{{\stackrel{0}{M}}_{jj} - 
{\stackrel{0}{M}}_{ii}} \;\; (ij = 14\,,\,25\,,\,36)\,.
\end{equation}

\ni Here, the second term is a perturbation in the small parameter 
$(s_{ij}/ c_{ij})^2$. Henceforth, we will accept this situation.

Now, let us make two conjectures on the neutrino mass matrix 
${\stackrel{0}{M}} = \left( {\stackrel{0}{M}}_{ij} \right)$ given 
explicitly in Eq. (14). Namely, ({\it i}) its diagonal elements are

$$
{\stackrel{0}{M}}_{11} = {\stackrel{0}{M}}_{22} = {\stackrel{0}{M}}_{33} = 
{\stackrel{0}{m}} \gg {\stackrel{0}{M}}_{44} = {\stackrel{0}{M}}_{55} = 
{\stackrel{0}{M}}_{66} = 0 
\eqno{(18{\rm a})}
$$

\ni or

\vspace{-0.3cm}

$$
{\stackrel{0}{M}}_{11} = {\stackrel{0}{M}}_{22} = {\stackrel{0}{M}}_{33} = 
0 \ll {\stackrel{0}{M}}_{44} = {\stackrel{0}{M}}_{55} = 
{\stackrel{0}{M}}_{66} = {\stackrel{0}{m}} \,,
\eqno{(18{\rm b})}
$$

\ni implying through Eq. (17) that

\vspace{-0.3cm}

$$
m_{i} \simeq {\stackrel{0}{m}} + \frac{{\stackrel{0}{M}} 
\,\!^2_{ij}}{{\stackrel{0}{m}}}\;\;, \;\; m_{j} \simeq  - 
\frac{{\stackrel{0}{M}} \,\!^2_{ij}}{{\stackrel{0}{m}}} \;\; 
(ij = 14\,,\,25\,,\,36)
\eqno{(19{\rm a})}
$$

\ni or

\vspace{-0.3cm}

$$
m_{i} \simeq - \frac{{\stackrel{0}{M}} \,\!^2_{ij}}{{\stackrel{0}{m}}}\;\;, 
\;\; m_{j} \simeq {\stackrel{0}{m}} + \frac{{\stackrel{0}{M}} 
\,\!^2_{ij}}{{\stackrel{0}{m}}} \;\; (ij = 14\,,\,25\,,\,36)\,,
\eqno{(19{\rm b})}
$$

\ni respectively. And, ({\it ii}) its off--diagonal elements, forming its 
Dirac component ${\stackrel{0}{M}}\, \!^{(D)} = {\rm diag} 
({\stackrel{0}{M}}_{14}\,,\, {\stackrel{0}{M}}_{25}\,,\, 
{\stackrel{0}{M}}_{36})$, are in their structure similar to the diagonal 
elements of charged--lepton and quark matrices of the universal form (6), thus

\addtocounter{equation}{+2}

\vspace{-0.3cm}

\begin{equation}
{\stackrel{0}{M}}_{14} = \frac{\mu}{29} \,\varepsilon \;,
\;{\stackrel{0}{M}}_{25} = \frac{\mu}{29}\, \frac{4}{9}\, 
(80 +\varepsilon) \;,\; {\stackrel{0}{M}}_{36} = \frac{\mu}{29}\, 
\frac{24}{25}\,(624 + \varepsilon) \;,
\end{equation}

\ni where we put approximately $\varepsilon = 0$ [already for charged 
leptons $\varepsilon $ is small, {\it cf.} Eq. (26)].

From Eqs. (19) [in the option (a) or (b)] and (20) we calculate (with 
$\varepsilon = 0$)

\vspace{-0.3cm}

$$
\Delta m^2_{21} \!=\! m^2_2 \!-\! m^2_1 \!=\! 2{\stackrel{0}{M}}\,\!^2_{25} 
\!=\! 3.00 \mu^2 \,,\, \Delta m^2_{32} \!=\! m^2_3 \!-\! m^2_2 \!=\! 
2({\stackrel{0}{M}}\,\!^2_{36} \!-\! {\stackrel{0}{M}}\,\!^2_{25})\!=\! 
8.50 \times 10^2 {\mu}^2
\eqno{(21{\rm a})}
$$

\ni or

\vspace{-0.3cm}

$$
\Delta m^2_{21} \!=\! m^2_2 \!-\! m^2_1 \!=\! 
\frac{{\stackrel{0}{M}}\,\!^4_{25}}{{\stackrel{0}{m}}\,\!^2} \!=\! 2.26  
\frac{{\mu}^4}{{\stackrel{0}{m}}\,\!^2} \,,\, \Delta m^2_{32} \!=\! 
m^2_3 \!-\! m^2_2 \!=\! \frac{{\stackrel{0}{M}}\!^4_{36} \!-\! 
{\stackrel{0}{M}}\!^4_{25}} {{\stackrel{0}{m}}\,\!^2} \!=\! 
1.82\times 10^5 \frac{{\mu}^4}{{\stackrel{0}{m}}\,\!^2} \,.
\eqno{(21{\rm b})}
$$

\ni Thus,

\vspace{-0.3cm}

$$
\frac{\Delta m^2_{32}}{\Delta m^2_{21}} = 2.83\times 10^2 \;\;{\rm or}\;\; 
8.06\times 10^4\,.
\eqno{(22{\rm a\; or\; b})}
$$

\addtocounter{equation}{+2}

\ni Hence, putting

\begin{equation} 
\Delta m^2_{32} \equiv \Delta m^2_{\rm atm} \sim (3.0 \;\; {\rm to} \;\; 
3.5)\times 10^{-3} {\rm eV}^2\,,
\end{equation} 

\ni as for atmospheric $\nu_\mu $'s in the Super--Kamiokande experiment, 
we {\it predict} for solar $\nu_e $'s [from Eq. (22) in the option (a) 
or (b)] that

\vspace{-0.3cm}

$$
\Delta m^2_{\rm sol} \equiv \Delta m^2_{21} \sim (1.1\;\;{\rm to}\;\;1.2)
\times 10^{-5} {\rm eV}^2 \;\; {\rm or}\;\; (3.7 \;\;{\rm to}\;\; 4.3)
\times 10^{-8}{\rm eV}^2 \,.
\eqno{(24{\rm a\; or\; b})}
$$

\ni So, the first option (a) is not inconsistent with the mass--square 
scale of Large Mixing Angle (LMA) solar solution, while the second option 
(b) is closer to the LOW solar solution \cite{W5}. The Super--Kamiokande 
estimate 
(23) determines also [from Eq. (21) in the option (a) or (b)] the constant

\vspace{-0.3cm}

$$
\mu^2 \sim (3.5 \;\;{\rm to}\;\; 4.1) \times 10^{-6}{\rm eV}^2 
\;\;{\rm or}\;\;(1.3\;\; {\rm to}\;\; 1.4)\times 10^{-4}
\,{\stackrel{0}{m}}\;{\rm eV}\;
\eqno{(25{\rm a\; or\; b})}
$$

\ni for neutrinos. This gives $\mu \sim \left( 1.9 \;\;{\rm to}\;\; 
2.0\right) \times 10^{-3} $ eV or $\left( 1.1\;\;{\rm to}\;\;1.2 \right) 
\times 10^{-2}\sqrt{ {\stackrel{0}{m}}\,{\rm eV}\, } $, respectively.

Note that for charged leptons we got previously \cite{W1,W2}

\addtocounter{equation}{+2}

\vspace{-0.3cm}

\begin{equation} 
\mu^{(e)} = 85.9924\,{\rm MeV}\;,\; \varepsilon^{(e)} = 0.172329\;,\; 
\left( \frac{\alpha^{(e)}}{ \mu^{(e)}} \right)^2 = 0.023^{+0.029}_{-0.025} \;,
\end{equation} 

\ni when we fitted precisely their three masses 
$ m_e\,,\, m_\mu\,,\, m_\tau $ using our mass matrix (6). The error limits 
for $(\alpha^{(e)}/ \mu^{(e)})^2 $ came out from the actual error limits 
for $ m_\tau = 1777.03^{+0.30}_{-0.26} $ MeV \cite{W6}. With 
$\alpha^{(e)} = 0$ 
we obtained $ m_\tau = 1776.80 $ MeV as a {\it prediction}, when we used 
the experimental values of $ m_e $ and $ m_\mu $ as inputs.

More generally, Eqs. (19) and (20) with the estimate (25) [in the option 
(a) or (b)] lead to the following neutrino mass spectrum:

\vspace{-0.3cm}

\begin{eqnarray*}
m_1 & = & {\stackrel{0}{m}}\;,\; m_2 \simeq {\stackrel{0}{m}} + 1.50 
\frac{\mu^2}{{\stackrel{0}{m}}} \sim {\stackrel{0}{m}} + 
\frac{(5.3\,{\rm to}\,6.2)\times 10^{-6}\,{\rm eV}^2}{\stackrel{0}{m}}\,, \\  
m_3 & \simeq & {\stackrel{0}{m}} + 427\frac{\mu^2}{{\stackrel{0}{m}}} 
\sim {\stackrel{0}{m}} + \frac{(1.5\,{\rm to}\,1.8)
\times 10^{-3}\,{\rm eV}^2}{\stackrel{0}{m}}
\end{eqnarray*}

\vspace{-1,5cm}

\begin{flushright}
(27a)
\end{flushright}
 
\ni or

\vspace{-0.3cm}

\begin{eqnarray*}
m_1 & = & \;0\,\;,\; m_2 \simeq - 1.50 \,\frac{\mu^2}{{\stackrel{0}{m}}} 
\sim - (1.9\,{\rm to}\,2.1)\times 10^{-4}\,{\rm eV}\,, \\  
m_3 & \simeq & - 427 \,\frac{\mu^2}{{\stackrel{0}{m}}} \sim - 
(5.5\,{\rm to}\,5.9)\times 10^{-2}\,{\rm eV}\,,
\end{eqnarray*}

\vspace{-1.5cm}

\begin{flushright}
(27b)
\end{flushright}
 
\ni and

\vspace{-0.3cm}
 
\begin{eqnarray*}
m_4 & = & 0\;,\; m_5 \simeq - 1.50\, \frac{\mu^2}{{\stackrel{0}{m}}} 
\sim - \frac{(5.3\,{\rm to}\,6.2)\times 
10^{-6}\,{\rm eV}^2}{\stackrel{0}{m}}\,, \\  
m_6 & \simeq & - 427\ \frac{\mu^2}{{\stackrel{0}{m}}} \sim - 
\frac{(1.5\,{\rm to}\,1.8)\times 10^{-3}\,{\rm eV}^2}{\stackrel{0}{m}}
\end{eqnarray*}

\vspace{-1.5cm}

\begin{flushright}
(28a)
\end{flushright}

\ni or

\vspace{-0.3cm}
 
\begin{eqnarray*}
m_4 & = & {\stackrel{0}{m}}\;,\; 
m_5 \simeq {\stackrel{0}{m}} + 1.50\, \frac{\mu^2}{{\stackrel{0}{m}}} 
\sim {\stackrel{0}{m}} + (1.9\,{\rm to}\,2.1)\times 10^{-4}\,{\rm eV}\,, \\  
m_6 & \simeq & {\stackrel{0}{m}} + 427 \,\frac{\mu^2}{{\stackrel{0}{m}}} 
\sim {\stackrel{0}{m}} + (5.5\,{\rm to}\,5.9)\times 10^{-2}\,{\rm eV}\,.
\end{eqnarray*}

\vspace{-1,5cm}

\begin{flushright}
(28b)
\end{flushright}

\ni Note that here $ m_1\,,\, m_2\,,\, m_3 \gg |m_4|\,,\, |m_5|\,,\, |m_6|$ 
or $|m_1|\,,\, |m_2|\,,\, |m_3| \ll m_4\,,\, m_5\,,\, m_6 $. Also the second 
of these two options differs essentially from the familiar seesaw mechanism, 
where $m_4\,,\, m_5\,,\, m_6 $ are of a very high mass scale determined by a 
Grand Unification Theory, while the mass scale $ \stackrel{0}{m}$ may be 
{\it e.g.} of the order of 1 eV. Of course, $\sum_i m_i = 3\stackrel{0}{m}$. 

Similarly, the mixing tangents (16) become 

\vspace{-0,3cm}
 
\begin{eqnarray*}
\frac{s_{14}}{c_{14}} & = & 0\;,\, \frac{s_{25}}{c_{25}} \simeq - 1.23 
\frac{\mu}{{\stackrel{0}{m}}} \sim - (2.3\,{\rm to}\,2.5)\times 
10^{-3}\,\frac{\rm eV}{\stackrel{0}{m}}\,, \\  
\frac{s_{36}}{c_{36}} & \simeq & - 20.7\frac{\mu}{{\stackrel{0}{m}}} 
\sim - (3.9\,{\rm to}\,4.2)\times 10^{-2}\,\frac{\rm eV}{\stackrel{0}{m}}
\end{eqnarray*}

\vspace{-1,7cm}

\begin{flushright}
(29a)
\end{flushright}

\ni or 

\vspace{-0.3cm}
 
\begin{eqnarray*}
\frac{s_{14}}{c_{14}} & = & 0\;,\, \frac{s_{25}}{c_{25}} \simeq 1.23 
\frac{\mu}{{\stackrel{0}{m}}} \sim (1.4\,{\rm to}\,1.4)\times 10^{-2}\,
\sqrt{\frac{\rm eV}{\stackrel{0}{m}}}\,, \\  
\frac{s_{36}}{c_{36}} & \simeq & 20.7\frac{\mu}{{\stackrel{0}{m}}} 
\sim (2.3\,{\rm to}\,2.4) \times 10^{-1}\,
\sqrt{\frac{\rm eV}{\stackrel{0}{m}} }\,.
\end{eqnarray*}

\vspace{-1,5cm}

\begin{flushright}
(29b)
\end{flushright}

\vspace{0.4cm}

\ni {\bf 3. Neutrino oscillations}

\vspace{0.2cm}

We start with the familiar formulae for probabilities of neutrino 
oscillations $\nu_\alpha \rightarrow \nu_\beta $ on the energy shell,

\addtocounter{equation}{+3}

\begin{equation} 
P(\nu_\alpha \rightarrow \nu_\beta) = |\langle \nu_\beta| 
e^{i PL} |\nu_\alpha \rangle |^2 = \delta _{\beta \alpha} - 
4\sum_{j>i} U^*_{\beta j} U_{\beta i} U_{\alpha j} U^*_{\alpha i} 
\sin^2 x_{ji} \;,
\end{equation}

\ni valid if the quartic products $U^*_{\beta j} U_{\beta i} 
U_{\alpha j} U^*_{\alpha i} $ are real, what is certainly true when 
a possible CP violation can be ignored [then $U^*_{\alpha i} \!=\! 
U_{\alpha i} $, as in our case of Eq. (11), and $P(\nu_\alpha \rightarrow 
\nu_\beta) \!=\! P(\nu_\beta \rightarrow \nu_\alpha) $].  In Eq. (30) 

\begin{equation} 
x_{ji} = 1.27 \frac{\Delta m^2_{ji} L}{E} \;,\; \Delta m^2_{ji}  = 
m^2_j - m^2_i \;,
\end{equation} 

\ni where $\Delta m^2_{ji}$, $L$ and $E$ are measured in eV$^2$, km 
and GeV, respectively ($L$ and $E$ denote the experimental baseline 
and neutrino energy, while $ p_i = \sqrt{E^2 - m_i^2} \simeq E -m^2_i/2E $ 
are eigenvalues of the neutrino momentum $P$).

If $m^2_1 \simeq m^2_2 \simeq m^2_3 \simeq \,\stackrel{0}{m}\!^2 $, where 
$\Delta m^2_{21} \ll \Delta m^2_{32}$, and $m^2_4 \simeq m^2_5 \simeq m^2_6 \ll \,\stackrel{0}{m}\!^2 $ as well as $ s^2_{ij} \ll c^2_{ij}\; (ij = 14\,,\,25\,,\,36)$, as is true in the case of $\mu^2 \ll \,\stackrel{0}{m}\!^2 $, then the oscillation formulae (30) give in particular

\begin{eqnarray} 
P(\nu_e \rightarrow \nu_e)_{\rm sol}\;\;\;\, & \simeq & 1 -  c^2_{25} 
\sin^2 (x _{21})_{\rm sol} - \frac{1}{2} (1+c_{25}^2) s^2_{25} \;, 
\nonumber \\
P(\nu_\mu \rightarrow \nu_\mu)_{\rm atm}\;\; & \simeq & 1 - 
\frac{1}{2}(1 + c^2_{25})c^2_{36} \sin^2 (x _{32})_{\rm atm} - 
\frac{1}{8}(1 + c^2_{25} + 2 c^2_{36})(s^2_{25} + 2 s^2_{36})\;, \nonumber \\
P(\nu_\mu \rightarrow \nu_e)_{\rm LSND} & \simeq & \frac{1}{2} 
s^4_{25}\sin^2 (x _{25})_{\rm LSND}\;, \nonumber \\
P(\bar{\nu}_e \rightarrow \bar{\nu}_e)_{\rm Chooz} & \simeq & 1 - 
(1 + c^2_{25})s^2_{25} \sin^2 (x _{25})_{\rm Chooz}
\end{eqnarray} 

\ni for solar $\nu_e$'s, atmospheric $\nu_\mu$'s, LSND $\nu_\mu$'s and 
Chooz $\bar{\nu}_e$'s. Here, $\Delta m^2_{21} \equiv \Delta m^2_{\rm sol} 
\sim 10^{-5}$ eV$^2 $ for LMA solar solution, $\Delta m^2_{32} \equiv 
\Delta m^2_{\rm atm} \sim (3\;{\rm to}\; 3.5)\times 10^{-3}$ eV$^2 $ from 
the Super--Kamiokande atmospheric experiment and $|\Delta m^2_{25}| \equiv 
\Delta m^2_{\rm LSND} \sim\; e.\,g.\; 1$ eV$^2 $ for the LSND experiment. 
This is consistent with our previous {\it identification} (23) of 
$\Delta m^2_{32}$ and {\it prediction} (24) for $\Delta m^2_{\rm sol}$. The 
first two Eqs. (32) differ, strictly speaking, from the familiar two--flavor 
oscillation formulae (used in analyses of solar experiments \cite{W5}) by some 
additive terms that, fortunately, are small enough when $\mu^2 \ll 
\,\stackrel{0}{m}\!^2 $. In fact, from Eqs. (29) and the estimate (25) 
[in the option (a) or (b)]

\begin{eqnarray*} 
\!\!\!s^2_{25} & \!\!\!\simeq\!\!\! & 1.50 \left( 
\frac{\mu}{\stackrel{0}{m}}\right)^2  c^2_{25} \sim 
(5.3\;{\rm to}\;6.2)\!\times\! 10^{-6} \left( 
\frac{\rm eV}{\stackrel{0}{m}} \right)^2{\rm or}\;\,(1.9\; 
{\rm to}\; 2.1)\!\times\! 10^{-4} \frac{\rm eV}{\stackrel{0}{m}}\,, \\
\!\!\!s^2_{36} & \!\!\!\simeq\!\!\! & 427 \left( 
\frac{\mu}{\stackrel{0}{m}}\right)^2 \! c^2_{36} 
\sim (1.5\;{\rm to}\; 1.8)\!\times\! 10^{-3} \left( 
\frac{\rm eV}{\stackrel{0}{m}} \right)^2{\rm or}\,
(5.5\,{\rm to}\, 5.9)\!\times\! 10^{-2}\frac{\rm eV}{\stackrel{0}{m}} \,,
\end{eqnarray*} 

\vspace{-1.62cm}

\begin{flushright}
(33\,a or b)
\end{flushright}

\vspace{0.5cm}

\ni  where $s^2_{25} \ll c^2_{25} \sim 1$ and $s^2_{36} 
\ll c^2_{36} \sim 1$ for $\mu^2 \ll \,\stackrel{0}{m}\!^2 $. 
Hence, in our texture, the solar and atmospheric oscillation 
amplitudes are practically maximal,

\addtocounter{equation}{+1}

\begin{equation} 
\sin^2 2\theta_{\rm sol} \equiv c^2_{25} \sim 1 \;\,,\,\; 
\sin^2 2\theta_{\rm atm} \equiv \frac{1}{2}(1 +c^2_{25}) c^2_{36} \sim 1 \,,
\end{equation} 

\ni when $\mu^2 \ll \,{\stackrel{0}{m}}\,\!^2 $. If in the matrix 
$ U^{(3)}$ given in Eq. (9) there were $c_{12} \stackrel{>}{\sim} 1/\sqrt{2}$ 
and $s_{12} \stackrel{<}{\sim} 1/\sqrt{2}$ instead of 
$c_{12} = 1/\sqrt{2} = s_{12} $, then $\sin^2 2\theta_{\rm sol} 
\equiv (2c_{12}s_{12})^2 c^2_{25}$ and $\sin^2 2\theta_{\rm atm} 
\equiv (s^2_{12}+c^2_{12} c^2_{25})  c^2_{36}$ would be suitably smaller.

From the third Eq. (32) we can see that in our texture the {\it predicted} 
LSND effect is potentially very small, perhaps unobservable, as having [due 
to Eq. (33) in the option (a) or(b)] the oscillation amplitude

$$
\sin^2 2\theta_{\rm LSND} \simeq \frac{1}{2} s^4_{25} \sim \left\{ 
\begin{array}{l} (1.4\;{\rm to}\;1.9) \times 10^{-11} \left( 
\frac{\rm eV}{\stackrel{0}{m}} \right)^4 \;\,{\rm or} \\ 
(1.9\;{\rm to}\;2.2) \times 10^{-8} \left( \frac{\rm eV}{\stackrel{0}{m}} 
\right)^2 \end{array}\right.\,,
\eqno(35{\rm a\;or \; b})
$$

\ni when $\mu^2 \ll \,{\stackrel{0}{m}}\,\!^2 $. If {\it e.g.} 
$\stackrel{0}{m} = O$(1 eV) $- \;O(10^{-2}$ eV), where still 
$\mu^2 \ll \,{\stackrel{0}{m}}\,\!^2$, then $\sin^2 2\theta_{\rm LSND} = 
O(10^{-11}) - O(10^{-3})$ or $O(10^{-8}) - O(10^{-4})$. The corresponding 
mass--square scale is

$$
\Delta m^2_{\rm LSND} \simeq |\Delta m^2_{25}| \simeq \left\{ 
\begin{array}{l} m^2_2 \simeq \,\stackrel{0}{m}\!^2 + (1.1\;{\rm to}\;1.2) 
\times 10^{-5}\, {\rm eV}^2 \;\,{\rm or} \\ m^2_5 \simeq 
\,\stackrel{0}{m}\!^2 + (3.9\;{\rm to}\;4.2) \times 10^{-4} 
{\stackrel{0}{m}}\, {\rm eV} \end{array}\right.\,,
\eqno(36{\rm a\; or\; b})
$$

\ni where ${\stackrel{0}{m}} = m_1$ or $m_4$, respectively.

The fourth Eq. (32) describes the Chooz experiment for reactor 
$\bar{\nu}_e$'s. Due to its negative result, $P(\bar{\nu}_e \rightarrow 
\bar{\nu}_e)_{\rm Chooz} \sim 1$, there appears the experimental constraint 
$(1 +c^2_{25})s^2_{25} \equiv \sin^2 2 \theta_{\rm Chooz} \stackrel{<}{\sim} 
0.1$ for $s^2_{25}$, if $\Delta m^2_{25} \equiv \Delta m^2_{\rm Chooz} 
\stackrel{>}{\sim} 0.1 \;{\rm eV}^2$ \cite{W7}. This implies for 
the LSND effect (in our texture) the very small Chooz upper bound  

\addtocounter{equation}{+2}

\begin{equation} 
\sin^2 2 \theta_{\rm LSND} \simeq \frac{1}{2} s^4_{25}  \stackrel{<}{\sim} 
1.3 \times 10^{-3}\,, 
\end{equation} 

\ni if $\Delta m^2_{25} \gg \Delta m^2_{32} \sim (3.0$ to $3.5)\times 10^{-3} 
\;{\rm eV}^2$, what is consistent with $\Delta m^2_{25} \stackrel{>}{\sim} 
0.1 \;{\rm eV}^2$ and gives $(x_{25})_{\rm Chooz} \gg (x_{32})_{\rm Chooz} 
\simeq (x_{32})_{\rm atm} = O(1)$ as $(x_{ji})_{\rm Chooz} \simeq 
(x_{ji})_{\rm atm}$ numerically. Then

\begin{equation} 
\sin^2(x_{25})_{\rm Chooz} \simeq \frac{1}{2} 
\end{equation} 

\ni in the fourth Eq. (32). When combined with the formula (35) [in the 
option (a) or (b)], the Chooz bound (37) leads to the lower limit

$$
\stackrel{0}{m}\, \stackrel{>}{\sim} (1.0\;{\rm to}\; 1.1) \times 10^{-2} 
\;{\rm eV}\;\,{\rm or}\;\, (3.8\;{\rm to}\;4.1) \times 10^{-3} \;{\rm eV}
\eqno{(39{\rm a\; or\; b})}
$$

\ni for ${\stackrel{0}{m}} = m_1$ or $m_4$, still consistent with our 
requirement $\mu^2 \ll \,\stackrel{0}{m}\!^2$. Thus, Eq. (36) [in the 
option (a) or (b)] gives the very small lower limit $\Delta m^2_{\rm LSND} 
\, \stackrel{>}{\sim} (1.1\;{\rm to}\;1.3) \times 10^{-4}{\rm eV}^2$ or 
$(1.6\;{\rm to}\; 1.8) \times 10^{-5} {\rm eV}^2$ for the mass--square 
scale $\Delta m^2_{\rm LSND}$.

Note that Eqs. (32) imply the sum rule

\addtocounter{equation}{+1}

\begin{equation} 
\sin^2 2 \theta_{\rm sol} + \sin^2 2 \theta_{\rm LSND} + \frac{1}{2} 
\sin^2 2\theta_{\rm Chooz} \equiv c^2_{25} + \frac{1}{2} s^4_{25}  + 
\frac{1}{2}(1 +  c^2_{25}) s^2_{25} = 1 
\end{equation} 

\ni for the solar, LSND and Chooz oscillation amplitudes. This leaves 
room for the LSND effect if $\sin^2 2 \theta_{\rm sol}  \equiv 
c^2_{25} < 1$ {\it i.e.}, if oscillations of solar $\nu_e$'s are 
not strictly maximal, what seems to be the case \cite{W5}.

\vspace{0.2cm}

{\bf 4.Conclusions}

\vspace{0.2cm}

We presented in this paper an effective texture for six Majorana 
conventional neutrinos (three active and three sterile), based on the 
$6\times 6$ mixing matrix defined in Eqs. (7) -- (10), and on the 
conjectures (18) [in the option (a) or (b)] and (20) for the resulting 
mass matrix. The conjecture (20) requires that the Dirac $3\times 3$ 
component of the unitarily transformed $6\times 6$ mass matrix 
$\stackrel{0}{M}= {\stackrel{1}{U}}\,\!^\dagger M \stackrel{1}{U}$ be 
{\it similar} in structure to the charged--lepton and quark $3\times 3$ 
mass matrices of the {\it universal form} (6), constructed previously 
with a considerable phenomenological success \cite{W1,W2}.

Such a texture {\it predicts} reasonably oscillations of solar $\nu_e$'s 
in the form not inconsistent with the LMA solar solution [for the 
option (a)], if the Super--Kamiokande value of the mass--square scale 
for atmospheric $\nu_\mu $'s is taken as an input. In both cases, neutrino 
oscillations are practically maximal. The proposed texture also 
{\it predicts} a potentially very small, perhaps unobservable, 
LSND effect with the oscillation amplitude of the order 
$O[10^{-11} (\,{\rm eV}/ {\stackrel{0}{m}} )^4]$ or 
$O[10^{-8} (\,{\rm eV}/ {\stackrel{0}{m}} )^2]$ and the mass--square scale 
of the order $O(\stackrel{0}{m}\!^2) + O(10^{-5} \,{\rm eV}^2)$ or 
$ O( \stackrel{0}{m} \!^2) + O(10^{-4} \stackrel{0}{m} \,{\rm eV})$ 
(for the option (a) or (b), respectively). The negative result of 
Chooz experiment imposes on the oscillation amplitude of LSND effect 
(in our texture) a very small upper bound of the order of $O(10^{-3})$, 
implying for $\stackrel{0}{m}$ a lower limit of the order 
$O(10^{-2} \,{\rm eV})$ or $O(10^{-3} \,{\rm eV})$. If {\it e.g.} 
$\stackrel{0}{m} = O(1 \:{\rm eV})$, then 
$\sin^2 2 \theta_{\rm LSND} = O(10^{-11})$ or $O(10^{-8})$ and 
$\Delta m^2_{\rm LSND} = O(1 \:{\rm eV}^2)$. In such a case, the LSND 
effect (in our texture) is, of course, unobservable (though it still 
exists in principle). Notice that the estimations following from the 
original LSND experiment \cite{W8} are {\it e.g.} 
$ \sin^2 2 \theta_{\rm LSND} = 
O(10^{-2})$ and $\Delta m^2_{\rm LSND} = O(1 \,{\rm eV}^2)$. The new 
miniBooNE experiment may confirm or revise the original LSND results.

Note finally that, as far as the neutrino mass spectrum is concerned, 
our model of neutrino texture is of 3 + 3, type in contrast to the models 
of 3 + 1 or 2 + 2 types \cite{W9} discussed in the case when, beside three 
active neutrinos $\nu_e\,,\,\nu_\mu\,,\,\nu_\tau $, one {\it extra} sterile 
neutrino $\nu_s $ exists. In these models, three Majorana conventional 
sterile neutrinos $\nu_{e_s}\,,\, \nu_{\mu_s} \,,\,\nu_{\tau_s}$ are 
decoupled through the familiar seesaw mechanism, as being practically 
identical with three very heavy mass states $\nu_4\,,\,\nu_5\,,\,\nu_6 $ 
(of the GUT mass scale). In our model $\nu_{e_s}\,,\, \nu_{\mu_s} 
\,,\,\nu_{\tau_s}$ are practically identical with three mass states 
$\nu_4\,,\,\nu_5\,,\,\nu_6 $  which [in the option (a)] are nearly 
massless or [in the option (b)] get masses only a little heavier 
({\it e.g.} of the order of 1 eV).

\end{document}